\documentclass[aps,pra,reprint,twocolumn,superscriptaddress,showpacs,showkeys,a4paper]{revtex4-1}

\usepackage{amsmath,amssymb,amstext}
\usepackage[usenames,dvipsnames]{color}
\usepackage{graphicx,subfigure,relsize}
\usepackage{bm,bbold,braket}
\usepackage{natbib}
\usepackage{umoline}

\definecolor{mygreen}{rgb}{0,0.5,0}
\definecolor{myblue}{rgb}{0,0,0.75}
\definecolor{mymagenta}{cmyk}{0,1,0,0.12}

\newcommand{\eq}[1]{\begin{equation} #1 \end{equation}}
\newcommand{\eqa}[1]{\begin{eqnarray} #1 \end{eqnarray}}

\newcommand{\vect}[1]{\ensuremath{\bm{#1}}}

\newcommand{\ue}{\mathrm{e}}

\newcommand{\LR}{Lieb--Robinson }

\begin{document}

%Title of paper
\title{Spread of correlations in long-range interacting quantum systems}

\author{P.~Hauke}
    \email{philipp.hauke@uibk.ac.at}    
    \affiliation{Institute for Quantum Optics and Quantum Information of the Austrian Academy of Sciences, A-6020 Innsbruck, Austria}
    \affiliation{Institute for Theoretical Physics, University of Innsbruck, A-6020 Innsbruck, Austria}
\author{L.~Tagliacozzo}
    \email{luca.tagliacozzo@icfo.es}
    \affiliation{ICFO-Institut de Ciencies Fotoniques, Av. Carl Friedrich Gauss, 3, 08860 Castelldefels, Barcelona, Spain.}

\date{\today}

\begin{abstract}

The non-equilibrium response of a quantum many-body system defines its fundamental transport properties and how initially localized quantum information spreads. 
However, for long-range-interacting quantum systems little is known. 
We address this issue by analyzing a local quantum quench in the long-range Ising model in a transverse field, where interactions decay as a variable power-law with distance $\propto r^{-\alpha}$, $\alpha>0$. 
Using complementary numerical and analytical techniques, we identify three dynamical regimes: short-range-like with an emerging light cone for $\alpha>2$; weakly long-range for $1<\alpha<2$ without a clear light cone but with a finite propagation speed of almost all excitations; and fully non-local for $\alpha<1$ with instantaneous transmission of correlations. This last regime breaks generalized Lieb--Robinson bounds and thus locality. 
Numerical calculation of the entanglement spectrum demonstrates that the usual picture of propagating quasi-particles remains valid, allowing an intuitive interpretation of our findings via divergences of quasi-particle velocities. 
Our results may be tested in state-of-the-art trapped-ion experiments.

\end{abstract}


%% 03.65.-w	Quantum mechanics
%   03.65.Ud	Entanglement and quantum nonlocality
%% 05.	Statistical physics, thermodynamics, and nonlinear dynamical systems
%   05.10.-a	Computational methods in statistical physics and nonlinear dynamics (see also 02.70.-c in mathematical methods in physics)
%   05.30.-d	Quantum statistical mechanics
%   05.45.-a	Nonlinear dynamics and chaos
%   05.70.Ln	Nonequilibrium and irreversible thermodynamics
%% 67.85.-d	Ultracold gases, trapped gases (see also 03.75.-b Matter waves in quantum mechanics) 
%   67.85.Bc	Static properties of condensates 
%   67.85.De	Dynamic properties of condensates; excitations, and superfluid flow
%% 75.	Magnetic properties and materials
%   75.10.Pq	Spin chain models
%   75.10.Jm Quantized spin models

% 03.65.Ud	Entanglement and quantum nonlocality
% 75.10.Pq	Spin chain models
% 67.85.-d	Ultracold gases, trapped gases
% 05.70.Ln	Nonequilibrium and irreversible thermodynamics


\maketitle

Physics is about identifying which in Nature are the causes and which are their effects. In abstract mathematical theories, however, this distinction is not always given. 
While special relativity was designed with the purpose of enforcing the causality principle, in non-relativistic quantum mechanics none of the five postulates ensures causality.
In that case, causality \emph{emerges} as a consequence of the locality of interactions. 
By now we have been able to see causality at work in well-controlled quantum-mechanical experiments described by local Hamiltonians, such as ultracold atoms \cite{Cheneau2012,Trotzky2012,Ronzheimer2013}. 
There, the spread of correlations is bounded by a light cone, similar to the spread of information in relativistic theories.
However, experiments are currently set up where quantum dynamics under variable long-range interactions can be studied, e.g., in polar molecules \cite{Lahaye2009,Carr2009,Trefzger2011}, Rydberg atoms \cite{Saffman2010,Loew2012}, or trapped ions \cite{Friedenauer2008,Kim2010,Islam2011,Lanyon2011,Islam2012,Britton2012}. 
This development makes it a pressing issue to answer the fundamental question: Can the out-of-equilibrium dynamics of synthetic long-range Hamiltonians effectively break causality?

We address this issue by studying a model that is currently realized in trapped-ion experiments, the transverse Ising model with long-range interactions. As we will show, the out-of-equilibrium response to an initially localized perturbation explores, depending on the interaction range, three different degrees of locality breaking. 
Specifically, we characterize the out-of-equilibrium response \cite{Polkovnikov2011} of the model to  \emph{local quenches}, obtained by perturbing locally the ground state of the system and observing its subsequent evolution. 

When the Hamiltonian that drives the evolution consists of local terms, the initially localized perturbation spreads at a finite speed, leading to the formation of a characteristic  `light cone' that bounds the propagation \cite{Cheneau2012}. This is a consequence of the  Lieb--Robinson bounds \cite{Lieb1972}, which in its essence formulates the principle of causality. Mathematically, under certain assumptions, the \LR bound expresses a bound for the time-dependent commutator between two operators ${\cal O}_A,   {\cal O}_B(t)$, defined at $t=0$ on two disjoint regions of the system  $A$ and $B$ separated by a distance $L$ \cite{Hastings2010a,nachtergaele_much_2011}, 
\begin{equation}
 [{\cal O}_A,   {\cal O}'(t)_B] \le ||{\cal O}_A|| \ ||{\cal O}'_B|| g(L)  \frac{v t}{ L} , \label{eq:lbr}
\end{equation}
where on the right hand side the norm is the operator norm, $v$ the \LR velocity, and $g(L)$ an exponentially decaying function.
This bound has proven essential for understanding  the complexity of quantum states \cite{Hastings2010a,nachtergaele_much_2011}, allowing to formulate several general theorems, e.g., connecting excitation gaps and decay of correlations \cite{hastings_area_2007,masanes_area_2009}.

In some systems, the \LR bound can be understood using an intuitive pseudo-particle picture \cite{Calabrese2005,calabrese_time_2006,calabrese_entanglement_2007}. 
This applies if the low-lying excitations can be obtained by populating (for translational invariant systems) different pseudo-particle momentum states, with the vacuum characterized by the absence of pseudo-particles. 
Then, the system responds to a local perturbation by emitting  pseudo-particles propagating at different speeds. The fastest particles, which define the causal cone, propagate at a speed that is often identified as the \LR velocity for that specific model.

Much less is known about how correlations spread in the presence of long-range interactions, although these become important in many different contexts. 
Namely, in local models where some of the constituents propagate much faster than the others, one can capture the effect of the fast constituents in an effective description of the slow ones involving a non-local interaction. 
A prime example is Quantum Electrodynamics, describing the contact interaction of charges with photons propagating at the speed of light. In the non-relativistic limit, where the charges move much slower than the light, the presence of photons can be encoded in a  long-range Coulomb potential between the charges.
Theories with long-range interactions can have over-extensive energies \cite{Mukamel2009,Campa2009} and are thus strongly non-local. In such circumstances, one would  expect that concepts like causality and the locality of quasi-particle excitations should be reconsidered.

The purpose of this manuscript is to address this issue using complementary analytical and numerical calculations. 
We find three qualitatively different dynamical regimes, with a break-down of \LR bounds for strong long-range interactions, and a weaker form of locality breaking that obeys the \LR bounds for intermediate interaction ranges. We are able to explain these regimes via the above-mentioned pseudo-particle picture. 
Finally, we discuss experimental regimes in trapped-ion setups where our findings can be observed.

For this purpose, we study the out-of-equilibrium dynamics generated by long-range interactions in the simplest possible scenario that can be implemented in trapped-ions experiments \cite{Porras2004a}, namely the long-range transverse Ising chain (LRTI)
\eq{
\label{eq:HLRTI}
H=\sum_{\braket{i,j}}\sin(\theta)\frac{\sigma_i^x \sigma_j^x}{\left|i-j\right|^{\alpha}}+\cos(\theta) \sum_i \sigma_i^z\,. 
}
Here, $\sigma$ denote the usual spin-1/2 Pauli matrices, and we set fundamental energy unit and lattice spacing to unity. 
We consider a finite chain of $L$ sites with open boundary conditions.  
The parameter $\alpha$ is varied within the  broad limits $3 \gtrsim\alpha\gtrsim 0$ that can be realized in the ion setups, allowing to tune from effectively short-range to strong long-range physics.
The parameter $\theta$ is varied in the range of anti-ferromagnetic interactions, $0 \le \theta \le \frac{\pi} 2$.
For any $\alpha 
> 0$, the system has two gapped phases, a $z$-polarized phase for small $\theta$, and a N\'eel-ordered phase for values of $\theta \simeq \pi/2$. The two phases are separated by a line of second-order phase transitions, whose universality class depends on $\alpha$ \cite{Koffel2012}.

Although the LRTI model does not obey the bound \eqref{eq:lbr}, which only holds for exponentially decaying Hamiltonians, one can still find a generalized \LR bound  \cite{Hastings2006a,Cramer2008,Hastings2010a} if the power-law interactions are `reproducing.' 
This condition, equivalent to a sufficiently fast decay, is fulfilled for $\alpha > 1$ (see supplemental material \cite{supp_long-range_dynamics}), and bounds decay of correlations by a power law governed by $\alpha$.

\begin{figure}
\centering
\includegraphics[width=\columnwidth]{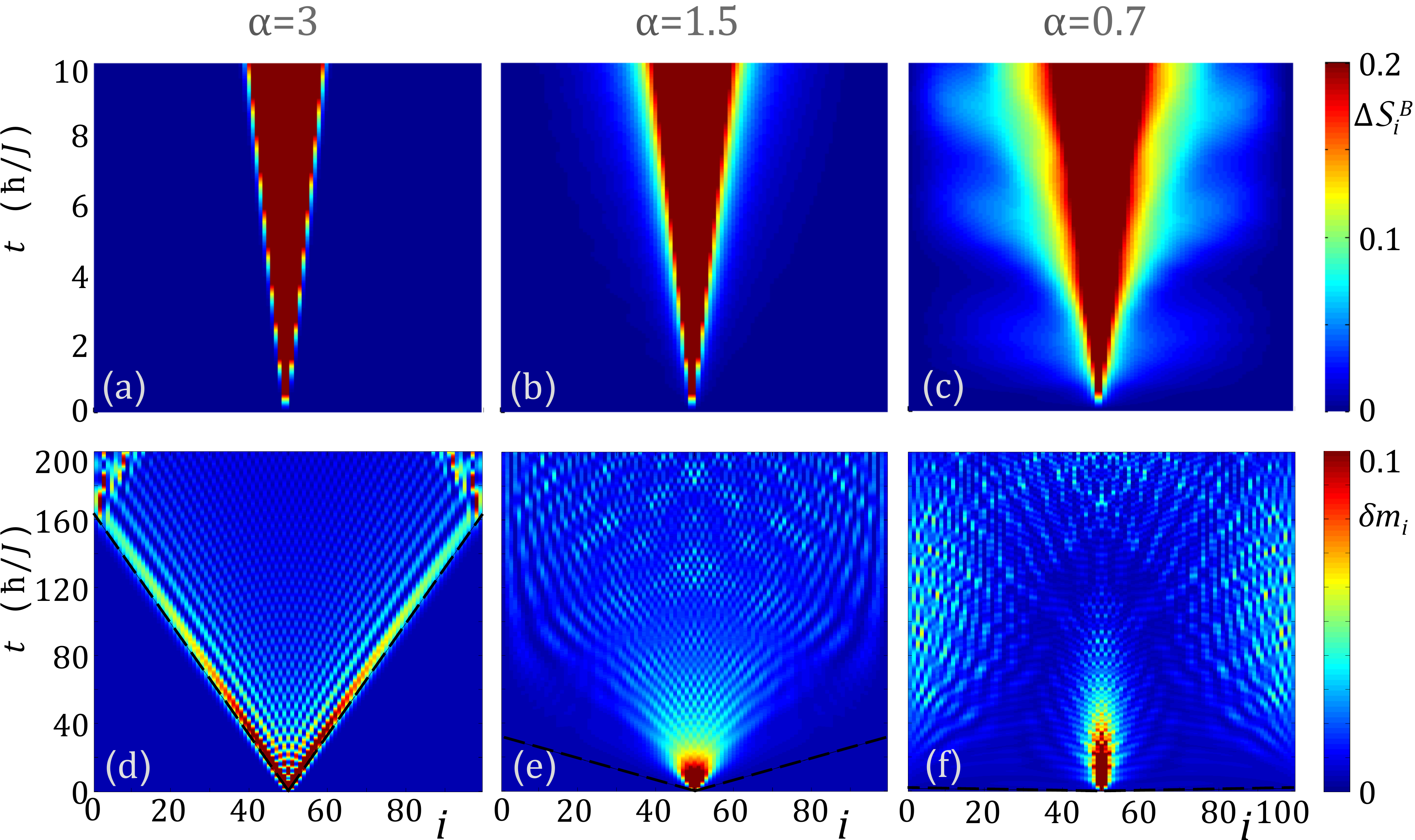}
\caption{ (Color online) {\bf (Non-)light cones.} 
{\bf (a-c)} Block entanglement entropy $\Delta \mathcal{S}_l = \mathcal{S}_l(t) - \mathcal{S}_l(0)$ from TDVP ($\theta=\pi/5$, $L=100$).
{\bf (d-f)} Polarization $\delta m_i=\braket{S_i^z}+1/2$ from LSWT ($\theta=\pi/20$). 
(a,d) For $\alpha>2$, the excitation at $i=50$ spreads light-cone like, as in the short-range model. 
(b,e) For $2>\alpha>1$, there is no well-defined wave front, but the excitation needs a finite time to bridge large distances. 
(c,f) For $\alpha<1$, the excitation spreads immediately over the entire system. 
Black dashed lines in (d-f) denote the maximal spin-wave group velocity [practically coinciding with the abscissa in (f)].
 }
\label{fig:lightcones}
\end{figure}

\emph{Numerical results---}
To study the effects of $\alpha$ on the out-of-equilibrium dynamics after a local quench, 
we use as initial state the ground state $\ket{\psi_{\mathrm{GS}}}$ of Hamiltonian \eqref{eq:HLRTI}  
at specific values of $\theta$ and $\alpha$, and at time $t=0$ perturb it locally; typically $\ket{\psi_0}=\sigma_{L/2}^x \ket{\psi_{\mathrm{GS}}}$. 
To observe the response of $\ket{\psi_{\mathrm{GS}}}$ to this local perturbation, we evolve $\ket{\psi_0}$ in time with the same Hamiltonian \eqref{eq:HLRTI}.

In our analysis, we employ two complementary approaches, the quasi-exact Time Dependent Variational Principle (TDVP) on matrix-product states (MPS) \cite{banuls_matrix_2009} and a linear spin-wave theory (LSWT) (see supplemental material \cite{supp_long-range_dynamics}). 
The used TDVP algorithm generalizes the ones available in the literature  \cite{crosswhite_applying_2008,mcculloch_infinite_2008,frowis_tensor_2010,haegeman_time-dependent_2011,Nebendahl2012,Koffel2012,milsted_variational_2012}. 
Here, we consider chain sizes up to $L=150$, and we have checked that the accuracy of MPS with matrix sizes $\chi\le 200$ is sufficient. 
The LSWT involves a higher degree of approximation, and is only valid for states with sufficient magnetic order. It has the advantage that it can access, with lower computational cost, larger times and system sizes than what is possible with the TDVP (we calculate numerically up to $L=1024$ and analytically for the thermodynamic limit). 
In those regimes where LSWT can be applied, we have checked  that the two methods provide compatible results, showing that the time evolution they describe is essentially semi-classical. 
This agreement is plausible, since $\ket{\psi_0}$ contains a single excitation with a density that decreases during the evolution, thus justifying the assumption of non-interacting quasi-particles that underlies the LSWT.

We exemplify the TDVP results for $\theta=\pi/5$ (Fig.~\ref{fig:lightcones}a-c), which is not accessible with LSWT because a nearby quantum phase transition strongly reduces magnetic order. 
We study the spread of quantum correlations via the block entanglement entropy (EE) $\mathcal{S}_l= -\sum_n \rho_l^n \log \rho_l^n $, where $\rho_l^n$ is the $n$-th eigenvalue of the reduced density matrix $\rho_l$ involving the spins $1,\dots, l$. 
As known from \cite{Koffel2012}, in the ground state of the $z$-polarized phase of the LRTI, the long-range interactions cause $\mathcal{S}_{L/2} \propto \log L$  when $\alpha<2$.
Therefore, to isolate the growth of the entropy generated during the time evolution, we analyze the excess of EE with respect to the initial state, $\Delta \mathcal{S}_l = \mathcal{S}_l(t) - \mathcal{S}_l(0)$.

For LSWT, we exemplify the resulting dynamics for $\theta=\pi/20$ (Fig.~\ref{fig:lightcones}d-f), 
where the ground state is strongly polarized, $\braket{S_i^{z}}\approx -1/2$ \cite{Koffel2012}. 
In this case, a useful measure for the spread of the perturbation is the excess magnetization  $\delta m_i=\braket{S_i^{z}}+1/2$. 
Notably, within LSWT, this directly gives the single-site entanglement entropy, $ {\mathcal{S}}_i^{(1)} = (\delta m_i +1) \log (\delta m_i + 1) - \delta m_i \log \delta m_i $ \cite{Peschel2009,Song2011}. 
Figure~\ref{fig:lightcones} evidences the similar behavior for the two methods and the two $\theta$ regimes.

For generic $\theta$, we identify three  dynamical regimes as a function of $\alpha$. 
(i) For $\alpha\geq 2$ [realized in Nature, e.g., for van-der-Waals ($\alpha=6$) or dipole--dipole ($\alpha=3$) interactions], the system behaves as if short-range interacting, with an excitation maximum that defines a clear wave front. Its linear propagation gives a constant \LR velocity, coinciding with the maximal spin-wave group velocity. 
Outside the resulting light cone, correlations decay algebraically with a power determined by $\alpha$, thus obeying the generalized \LR bounds. 
(ii) In the range $2>\alpha>1$, although at short times there appears an effect resembling a light cone, it does not really bound the propagation of the perturbation, since correlations consistently leak out of it, and at larger times one cannot identify a wave front. 
Further, we find complex interference effects due to longer-range spin flips. 
Still the excitation needs a finite time to bridge larger distances. 
(iii) For $\alpha<1$ ($\alpha=1$ corresponds to Coulomb- or gravitation-like potentials), 
the generalized \LR bounds valid for $\alpha>1$ can no longer be defined. Consequently, the system becomes truly long ranged, and correlations spread practically instantaneously over the chain. 

These results complement the one of \cite{Ziraldo2012} about thermalization in disordered systems, where random interactions are modulated by a long-range power law. 
There, the time average of local observables tends to a value predicted by a Generalized Gibbs ensemble only if $\alpha<1$. 

Our findings differ from previous results for the specific cases of Hamiltonians consisting of mutually commuting terms, such as Eq.~\eqref{eq:HLRTI} with $\theta=\pi/2$. In such settings, the block entropy of subsystems can increase unchecked with block size for $\alpha\leq 0.5$, whereas for $\alpha>1$ it is strictly upper bounded \cite{Duer2005}. Further, the value $\alpha=0.5$ separates two dynamical regimes \cite{Bachelard2013}, one of which is characterized by prethermalization plateaus \cite{vandenWorm2012}.

\begin{figure}
\centering
\includegraphics[width=1\columnwidth]{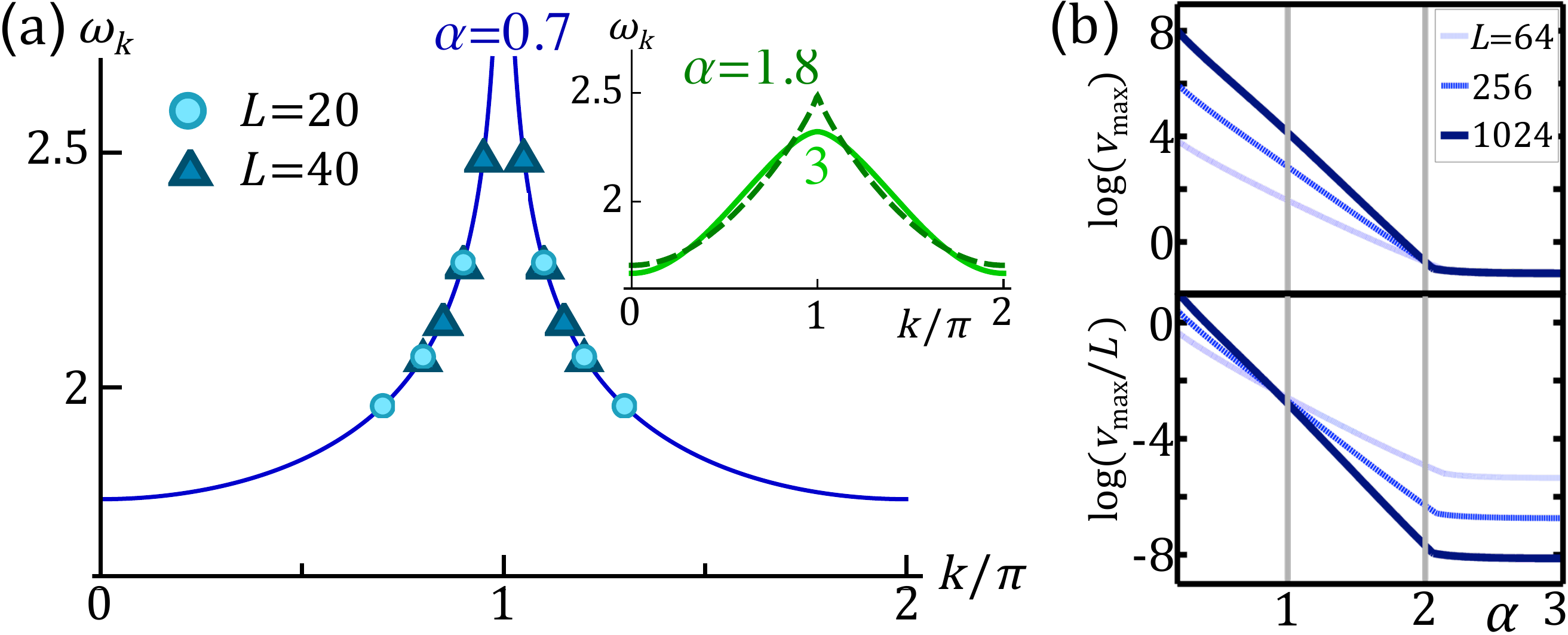}
\caption{ (Color online) {\bf (a) Spin-wave dispersion relations.} 
{\bf Inset:} For $\alpha>2$, $\omega_k$ is a deformed cosine, similar to the short-range case, while at $\alpha<2$, it develops a cusp at $k=\pi$, which becomes sharper with decreasing $\alpha$. 
{\bf Main panel:} 
For $\alpha<1$, the number of modes with diverging group velocity, $\left|v_g\right| > (L/2)/t_{0}$, increases with $L$ for any $t_{0}>0$. 
Plotted for $t_{0}=50$, with $L=20$ (6 modes, circles) and $L=40$ (8 modes, triangles).
{\bf (b) Maximal group velocity} for different $L$ (at $\theta=\pi/20$).
{\bf Top:} For $\alpha>2$, $v_{\mathrm{max}}$ is essentially independent of $\alpha$, while it increases sharply below $\alpha= 2$. 
{\bf Bottom:} For $\alpha>1$, $v_{\mathrm{max}}/L$ tends to zero for $L\to\infty$. The time $t_b\equiv L/(2 v_{\mathrm{max}})$ at which excitations reach the system boundary diverges. 
For $\alpha< 1$, $v_{\mathrm{max}}/L$ \emph{increases} with system size. Information about the local quench reaches the entire system instantaneously. 
}
\label{fig:dispersionrelations_analytical}
\end{figure}

\emph{Pseudo-particle dispersion relation---}
The qualitatively different behavior in the regimes (i-iii) can be understood in a simple quasi-particle picture: 
During the local quench, all spin-wave $k$-modes become populated with occupation $\approx 1/L$. 
If the pseudo-particles do not interact (a good approximation for low pseudo-particle density), each mode subsequently propagates with its group velocity $v_g=\frac{\partial\omega_k}{\partial k}$, which depends only on the dispersion relation $\omega_k$
(c.f.\ Fig.~\ref{fig:dispersionrelations_analytical}a; see \cite{supp_long-range_dynamics} for an analytical formula from LSWT). 

In the range $2<\alpha<\infty$, the maximal group velocity $v_{\mathrm max}$ is achieved around $k=\pi$, and does barely depend on system size or $\alpha$ (Fig.~\ref{fig:dispersionrelations_analytical}b, top). 
At $\alpha<2$, however, $\omega_k$ acquires a cusp at $k=\pi$. 
Consequently, $v_{\mathrm max}$ is attained at $k=\pi\pm2\pi/L$  \cite{Note2}. It diverges as $v_{\mathrm max}\propto (2\pi/L)^{\alpha-2}$. 
Still, the time scale in which pseudo-particles can reach the boundary, $t_b\equiv L/(2 v_{\mathrm max})$, scales as $L^{\alpha-1}$, which diverges for $1<\alpha\leq2$; the time to reach the boundary increases with system size, even for the fastest mode. 

The long-range effects become more dramatic at $\alpha<1$ due to a stronger divergence $v_{\mathrm{max}}\propto (2\pi/L)^{(\alpha-3)/2}$. Now, for the fastest mode, $t_b$ \emph{decreases} with system size (actually for a diverging number of modes, see Fig.~\ref{fig:dispersionrelations_analytical} and \cite{supp_long-range_dynamics}). 
In Fig.~\ref{fig:dispersionrelations_analytical}b, the transition between the three regimes can be clearly identified. 

The spin-wave dispersion also explains the diffusive effect encountered at small $\alpha$ (see Fig.~\ref{fig:lightcones}e-f). 
With decreasing $\alpha$, the dispersion becomes flatter around the sides of the Brillouin zone. Therefore, there are many slow quasi-particles that remain in the central region for a long time, giving rise to an apparent diffusive core of high density. 

\begin{figure}
\includegraphics[width=1\columnwidth]{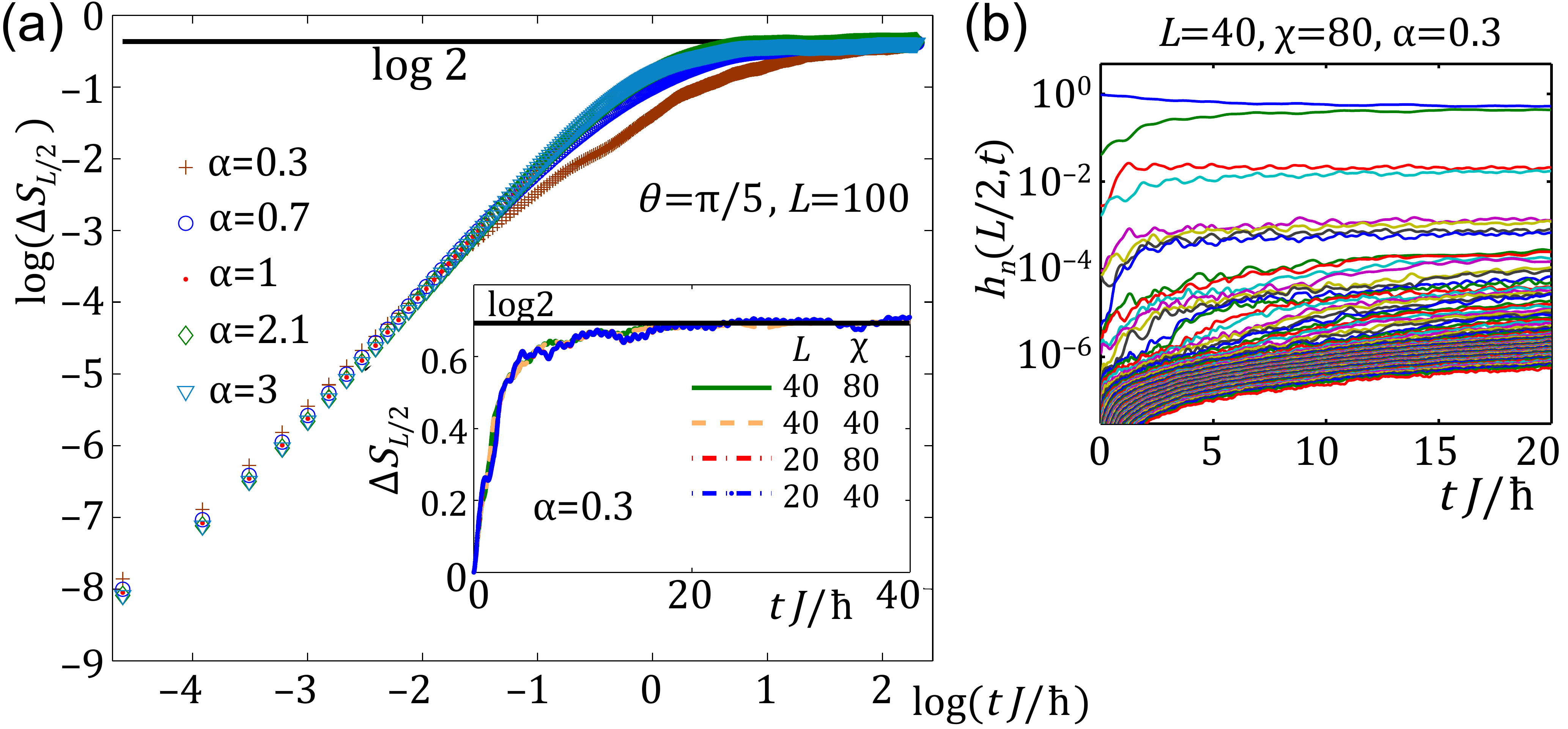}
\caption{ \label{fig:ent}(Color online) {\bf (a) Growth of entanglement entropy.} The excess $\Delta \mathcal{S}_{L/2}(t)$ grows initially as a power law with $t$ for all considered $\alpha$. It then saturates to $\log 2$ independent of system size, as expected from the pseudo-particle picture. This is shown in the insets where we compare the saturation value for chains of different length and for completeness show that there is no residual dependence of the saturation value on the  MPS matrix dimension $\chi$.
{\bf (b) Evolution of the entanglement spectrum.} The entanglement spectrum is dominated by two eigenvalues, which in the pseudo-particle picture correspond to the pseudo-particle being in the left or the right part of the chain. The other eigenvalues are significantly smaller, confirmating the quality of semi-classical descriptions of the evolution.}
\end{figure}

\emph{Scaling of entanglement entropy---}
To numerically confirm the validity of the pseudo-particle picture, we analyze within the TDVP the increase of the EE of half of the chain $\mathcal{S}_{L/2}(t)$.  
Interestingly, for all values of $\alpha$ considered, the excess entropy $\Delta  \mathcal{S}_{L/2}(t)$ initially increases as a power of $t$ and then saturates to a value very close to  $\Delta  \mathcal{S}_{L/2}(t)=\log 2 $, independent of system size (Fig.~\ref{fig:ent}a). 

The initial growth is faster for smaller $\alpha$, in agreement with the presence of faster pseudo-particles.
Remarkably, due to these fast pseudo-particles the initial growth is stronger than logarithmic which normally is considered the worst-case scenario, occuring at quenches to a critical point. 
Before entering the saturation regime, systems with smaller $\alpha$ start to evolve slower, in agreement with the appearance of a diffusive evolution. 
The fact that the excess of  EE of a block saturates to  a value independent of its size is in remarkable contrast to the ground-state properties. This effect finds a natural explanation in the semi-classical picture of pseudo-particles: the states that dominate the time evolution are states with only one pseudo-particle; the $\log 2$ is then immediately understood as coming from the two orthogonal possibilities of the pseudo-particle being either in the left or in the right half-chain. 

A further confirmation comes from the half-chain entanglement-spectrum evolution, $h_n(L/2,t) =\log \rho^n_{L/2}(t)$, where $\rho^n_{L/2}$ is the $n$-th eigenvalue of the reduced density matrix of half of the chain. The spectrum is dominated by only few eigenvalues, with two of order one as expected from the $\log 2$ asymptote, and a huge number of eigenvalues below $10^{-5}$ (Fig.~\ref{fig:ent}b). These eigenvalues grow steadily, but we expect that they do not affect equilibrium properties, since they are associated to higher energies and thus, at long times, their effect should average out. These findings are in agreement with similar observations in short-range systems \cite{Calabrese2005,calabrese_entanglement_2007,Eisler2007,Eisler2008,perales_entanglement_2008,Laeuchli2008,Fagotti2008,Igloi2009,rieger_semiclassical_2011,Stephan2011,Igloi2012}, where semi-classical models provided a good description of these kinds of out-of-equilibrium dynamics.

\emph{Experimental implementation---}
Due to their qualitative difference, the three dynamical regimes can be observed already in small experimental systems. 
A clear signature is, e.g., the speed with which excitations reach the boundary and its scaling with system size. 
Alternatively, the ratio of the wave-front maximum and the subsequent minimum distinguishes the short-range and the weakly long-range regime. In the former, it increases with system size, thus defining an increasingly sharp wave front. In the latter, it decreases until the wave front  disappears. 

Finally, let us remark that although the abstract LRTI model displays non-local behaviour, an actual physical implementation will obey locality, as one would expect. 
E.g., in the trapped-ion implementation, Hamiltonian \eqref{eq:HLRTI} describes an effective dynamics for electronic states of the ions, which are coupled by collective phonon modes by employing laser fields \cite{Porras2004a,Porras2006c,Schneider2012}. 
The phonon dynamics can be neglected on time scales much larger than those associated to the detuning between laser driving and phonon frequencies. These time scales are typically $\mathcal{O}(10\mu\mathrm{s})$. 
Moreover, the derivation of Eq.~\eqref{eq:HLRTI} employs a rotating-wave approximation in the phonon frequencies, corresponding to neglecting terms that average to zero on time scales $\mathcal{O}(1\mu\mathrm{s})$.  
When the group velocity reaches these time scales, the effective Hamiltonian  \eqref{eq:HLRTI} breaks down, just as how the Coulomb potential is no longer valid when charged particles move close to the speed of light. 
On the other hand, the time scale of the spin interactions is typically $\hbar/J=\mathcal{O}(1\mathrm{ms})$. 
Therefore, although the group velocities of the spin system cannot truly diverge, they can be several times larger than the scale set by $J$. 
This still provides a drastic effect that can be explored in typical practical implementations \cite{Friedenauer2008,Kim2010,Islam2011,Lanyon2011,Islam2012,Britton2012}.

\emph{Conclusions---}
Via quasi-exact numerics based on Tensor Networks and analytical calculations of the spin-wave dispersion, we have identified three qualitatively different regimes of the non-equilibrium dynamics in the LRTI model, indicating different degrees of the break-down of locality. 
The quasi-particle dispersion undergoes drastic changes at $\alpha=2$ and $\alpha=1$, marking a transition from short-range over weakly long-range to strong long-range physics. 
In the last case, diverging quasi-particle velocities lead to a practically instantaneous spread of excitations through the entire system. 
It will be interesting to study how these findings carry over to larger dimensions. 
Finally, we have outlined how to identify the different degrees of non-locality in typical trapped-ion experiments, and we hope our findings to inspire experiments along these lines.

Identifying violations of the \LR bounds -- besides establishing the presence/absence of causality in systems with long-range interactions -- may pave the way for extending well-established results about the complexity of ground states \cite{Hastings2010a,nachtergaele_much_2011}, and the relation between the decay of correlations and the scaling of entanglement \cite{hastings_area_2007,masanes_area_2009}. Moreover, the \LR bound has important implications for thermalization \cite{kinoshita_quantum_2006,rigol_relaxation_2007}: if the system locally equilibrates to a Generalized Gibbs ensemble, time-dependent correlation functions are described by the same ensemble \cite{essler_dynamical_2012}. 
These are key issues that have strong technical consequences for our ability to simulate the quantum system on a computer. 
Indeed, simulations based on Tensor Networks such as MPS, or the Multi-Scale Entanglement Renormalization Ansatz (MERA), typically require a small amount of entanglement --- but due to Lieb--Robinson bounds, correlations build up linearly during time evolution, making numerical simulations often unfeasible \cite{Note1}.

We acknowledge interesting discussions with P.\ Calabrese, J.\ Eisert, Th.\ Koffel, C.\ Laflamme, B.\ Lanyon, M.\ Lewenstein, F.\ Verstraete, and P.\ Zoller, and financial support from the Marie Curie project FP7-PEOPLE-2010-IIF ENGAGES 273524, TOQATA  (FIS2008-00784), Spanish MICINN (FIS2008-00784), Catalunya-Caixa, the Austrian Science Fund (SFB F40 FOQUS), EU IP SIQS, and the DARPA OLE program. 

\emph{Note added.---}
During the review process of this article, two related preprints appeared, one that studies the LRTI under global quenches \cite{Schachenmayer2013}, and one that studies local quenches in a LRTI with interactions modelled after a realistic trapped-ion string with non-uniform inter-ion distances \cite{Gong2013}. Both obtain a transition of dynamical behavior \cite{schutzhold_dynamical_2008,heyl_dynamical_2013} at the same value $\alpha=1$ as this work.

%\appendix

\renewcommand{\theequation}{S\arabic{equation}}
\setcounter{equation}{0}
\renewcommand{\thefigure}{S\arabic{figure}}
\setcounter{figure}{0}

\vspace*{1cm}
{\begin{center}{\LARGE{\bf Supplemental material}}\end{center}}

In this supplemental material, we explain technical details of the TDVP algorithm and the time evolution in linear spin-wave theory. 
We also provide a proof that interactions decaying with a power-law are reproducing if and only if $\alpha>1$.

\section{Time-reversal-symmetric scheme for the TDVP applied to long-range Hamiltonians}

In a previous work, it has been shown how to find an MPS approximation to the ground state of a long-range Hamiltonian using an extension of the   time-dependent variational principle (TDVP) \cite{Koffel2012}.
The TDVP is an algorithm that uses the geometric notion of the MPS tangent plane \cite{haegeman_time-dependent_2011,milsted_variational_2012}. 
It allows to find the ground-state description as a MPS by solving a differential equation for the tensors defining the MPS.
The same generalization of the TDVP presented in \cite{Koffel2012}  can be used to perform real-time evolutions \cite{haegeman_time-dependent_2011,milsted_variational_2012} for systems with long-range interaction, which we have exploited to compute the quench dynamics described in the main text.
The main difference when performing the dynamics -- instead of imaginary time as is necessary to obtain the ground state -- in real time, is that special care has to be taken to ensure that the algorithm does not violate time-reversal symmetry. This immediately ensures that the algorithm  conserves both the norm and the energy of the initial state, as it should. 
Care is advised, as real-time evolution does not enjoy the same self-correction as imaginary-time evolutions where small errors in one step can be corrected in the next step.
Here, instead, any small error is propagated along the simulation, and one needs further care to minimize those.

Here, we describe the steps we follow in order to ensure the time-reversal invariance of the integrators scheme that we apply to the above-mentioned differential equations. 
This is a simple modification of the technique proposed in   \cite{haegeman_time-dependent_2011,verschelde_variational_2011}, suitable for evolutions dictated by long-range Hamiltonians in finite chains.
 
First, we briefly recall the general strategy that is common to both real-time and imaginary-time dynamics.
 
{\sl (1)} Encode the starting state of the time evolution as a MPS  described by the set of tensors $L$ $\set{A_n}$, $A_n, n=1\cdots L$ $\ket{\psi_0\{A\}}$  and {\sl (2)} encode the long-range Hamiltonian as a MPO described by a set of tensors $L$  $\set{O}$, $H({\set{O}})$.  The evolved state is obtained by {\sl (3)} solving the Schr{\"o}dinger equation for a short time interval $ dt$ with initial condition given by $\ket{\psi_0}$, 
 \begin{equation}
i \partial_t \ket{\psi(\{A(t)\})} =   H(\set{O}) \ket{\psi(\{A(t)\})}\,, \label{eq:Schr}
\end{equation}
  where we set $\hbar=1$.
 {\sl (4)} To solve the above equation inside the manifold of MPS with fixed tensor dimensions, one needs to introduce tangent vectors.
 These are generically defined through two sets of tensors $\set{A}$ and $\set{B}$, and are expressed as the linear combination of MPS defined by $A$ everywhere but one $B_n$ at a specific place, in formula
 \begin{equation}
  \ket{T(\set{A},\set{B})} = \sum_n \ket{\psi(\{A\}_n, B_n)}\,,
 \end{equation}
 where we have used the notation $\set{A}_n$ to define the set of $A$ tensors from which we have removed tensor $A_n$.
 
{\sl (5)} While the left hand side of the above equation \eqref{eq:Schr} defines a tangent space to the manifold of the MPS with fixed bond dimension, the right hand side is not contained in that space and should be explicitly projected onto it. In formula, we would like to find the tangent vector $\ket{T}$ that minimizes the distance from  $H \ket{\psi(\{A(t)\})}$,
\begin{equation}
 \ket{T{\set{A},\set{B^*}}}, : \min_{\ket{T}} || \ket{T} - H \ket{\psi(\{A(t)\})}||^2.
\end{equation}

In practice, in the canonical form, the computation is simplified by requiring that the tangent vectors are orthogonal to the original vector. To ensure the orthogonality, the $B_n$ tensors in the tangent vectors are defined as the contraction of auxiliary tensors, (for normalization convenience) the inverse square root of the reduced density matrix, times a matrix of free coefficients of dimension  called  $X_n$, and a fixed projector $V_n$ on the orthogonal space to the one on which the starting vector is defined.

{\sl (6)} At this point, one can discretize Eq.~\eqref{eq:Schr} and integrate  it iteratively through
\begin{equation}
 A_n(t+dt) = A_n(t) + i\, dt\,  (B_n)^*, \label{eq:single_step}
\end{equation}
for all sites  $n= 1\dots L$.

We now turn to real-time dynamics, and we focus specifically on designing a time-reversal-invariant integrator scheme. This requires improving the first-order integrator \eqref{eq:single_step} to at least the so called middle-point integrator. 
This involves finding an intermediate step for each $n$, $A_n(t + dt/2)$ such that 
both $A_n(t+dt) = A_n(t+ dt/ 2) + i dt/2  (\tilde B_n)^*$ and $A_n(t) = A_n(t+ dt/ 2) - i dt/2  ( \tilde B_n)^*$, where $\set{ \tilde B}$ is the set of tensors defining
the projection onto the tangent space of the action of the Hamiltonian on  the state $\ket{\psi(\set{A(t+dt/2)})}$. The two above conditions can be taken as a definition of $A_n(t + dt / 2)$ that we then use to complete the evolution step by just integrating the state for another $dt/2$, 
\begin{equation}
A_n(t+dt) = A_n(t+dt/2) + i dt/2  (\tilde B_n)^*,  \label{eq:mid_step}
\end{equation}
so that we ensure that the evolution is invariant under time reversal. 
The important part becomes finding the intermediate $A_n(t + dt / 2)$. As suggested in \cite{haegeman_time-dependent_2011}, one can devise an iterative procedure to determine  $A_n(t + dt / 2)$. 
Here, we describe an alternative procedure to the one presented in \cite{haegeman_time-dependent_2011} that is well suited for finite-chain Hamiltonians encoded in MPO as the ones discussed in this paper.
The procedure consists in proceeding locally in the evolution (site by site) requiring that each local step is time-reversal invariant.
For this reason, chosen a position $n$ in the chain, one proceeds by  

{\sl (1)} Obtaining a trial $A^0_n( t + dt/ 2)$ by solving Eq.~\eqref{eq:Schr} for a time step $ dt/2$, with the initial state locally described by $A_n(t)$. 
{\sl (2)} Obtaining  a trial $\tilde B$ by finding the best tangent vector that approximates the r.h.s.\ of Eq.~\eqref{eq:Schr}. 
{\sl (3)} Evolving back  $A^0_n( t + dt/ 2)$ to $\bar{A}^0_n(t)$ (that initially will differ from $A_n(t)$) by solving Eq.~\eqref{eq:Schr}  for a time step $ -dt/2 $, with an initial state locally described by $A_n(t+dt/2)$.
{\sl (4)} Compute the $\Delta A^0_i(t) = A_i(t)- \bar{A}^0_i(t) $ and project it onto the tangent space defined at  $A^0_i( t + dt/ 2)$. 
{\sl (5)} Compute the error as $ E^0_i=\sqrt{|| \ket{\psi \left( \set{A}_i, \Delta A^0_i(t)  \right) } ||}$. 
{\sl (6)} In this way, we can obtain the improved estimate of the middle point $A_i(t +dt/2)$ as  $A^1_i( t + dt/ 2) =A^0_i( t + dt/ 2) + P_{\tilde B^0} \Delta A^0_i(t)$, where $P_{\tilde B^0}$ is the projection on the tangent plane at the old estimate $A^0_n( t + dt/ 2)$.  

We then repeat the procedure  starting again from step {\sl (2)} and iterate as often as necessary in order  to bring the error in the inversion $E_n$   below the required precision (typically around $10^{-12}$). The procedure is repeated  for all sites, and at the end of a sweep from 1 to $L$, one completes an elementary evolution step of $dt$.
For more details about the other aspects of the algorithm and possible improvement using higher-order  Ruge--Kutta integration schemes, we refer the reader to the literature on the subject \cite{haegeman_time-dependent_2011,milsted_variational_2012, Koffel2012}.

\section{Linear spin-wave theory for long-range models}

To gain some analytical understanding of the dynamics of the long-range system described by Hamiltonian \eqref{eq:HLRTI}, we employ a linear spin-wave theory (LSWT). 
This theory is well known to yield good qualitative results in phases with strong magnetic order \cite{Diep2004}, 
such as the strongly $z$-polarized phase occurring for small $\theta$ \cite{Koffel2012}. In our numerical analysis, we will therefore focus on that case, although we will keep our derivations general.  
An advantage of LSWT is that the long-range interactions are implemented into the formalism without additional complications, as we will sketch now. 

\subsection{Determining the ground state}

As a first step to finding the ground state of spin waves, it is convenient to rotate the spins into a local, twisted coordinate system $x^{\prime},y^{\prime},z^{\prime}$, so that the new $z'$ axis is aligned with the quantization axis. 
In the antiferromagnetic case of $\theta>0$, a convenient form is to rotate the spin 1/2 operators into $\vect{S}_i^{\prime}={\mathcal R}_i \vect{S}_i$, where
\eq{
{\mathcal R}_i = 
\left( \begin{array}{ccc}
(-1)^i \cos\gamma & 0 & -\sin\gamma \\
0 & (-1)^{i+1} & 0 \\
(-1)^{i+1}\sin\gamma & 0 & -\cos\gamma \end{array} \right) \,.
}
Since only $S^x$ and $S^z$ operators occur in Hamiltonian \eqref{eq:HLRTI}, it is sufficient to restrict the rotation to the $xy$ plane. 
We keep the angle $\gamma$ free at this stage and will find it later through the minimum of the energy. 

\begin{widetext}
In terms of the rotated spin operators, the system Hamiltonian reads 
%\eqa{
%\label{eq:Hrotated}
%H&=&4\sin\theta \sum_{\braket{ij}}\frac{1}{\left|i-j\right|^\alpha} 
%\Bigl[ (-1)^{i+j} \cos^2\gamma S_i^{x \prime}  S_j^{x \prime}\nonumber \\
%& & + (-1)^{i+j+1} \sin\gamma \cos\gamma \bigl( S_i^{x \prime}  S_j^{z \prime} + S_i^{z \prime}  S_j^{x \prime} \bigr) \\
%& & +(-1)^{i+j}\sin^2\gamma S_i^{z \prime}  S_j^{z \prime} \Bigr] \nonumber \\
%& & - 2\cos\theta \sum_i \left( \sin \gamma S_i^{x \prime}  + \cos\gamma S_i^{z \prime} \right) \nonumber \,.
%}
\eqa{
\label{eq:Hrotated}
H&=&4\sin\theta \sum_{\braket{ij}}\frac{1}{\left|i-j\right|^\alpha} 
\Bigl[ (-1)^{i+j} \cos^2\gamma\, S_i^{x \prime}  S_j^{x \prime} + (-1)^{i+j+1} \sin\gamma \cos\gamma \bigl( S_i^{x \prime}  S_j^{z \prime} + S_i^{z \prime}  S_j^{x \prime} \bigr) 
+(-1)^{i+j}\sin^2\gamma S_i^{z \prime}  S_j^{z \prime} \Bigr] \nonumber \\
&- & 2\cos\theta \sum_i \left( \sin \gamma S_i^{x \prime}  + \cos\gamma S_i^{z \prime} \right)  \,.
}
\end{widetext}

For a state that is strongly polarized along the $z^{\prime}$ axis, one can approximate spin-$S$ operators (here $S=1/2$) by bosonic operators via the Holstein--Primakoff transformation \cite{Diep2004}, 
$S_i^{z \prime} \to S - a_i^\dagger a_i$, $S^+\to \sqrt{2S} a_i^\dagger \sqrt{1-\frac{a_i^\dagger a_i}{2S}}$, and $S^-\to \sqrt{2S}\sqrt{1-\frac{a_i^\dagger a_i}{2S}} a_i$. 
We now insert these into Hamiltonian~\eqref{eq:Hrotated}, neglect contributions beyond linear order in $\frac{a_i^\dagger a_i}{2S}$, and use that terms that are linear in the boson operators vanish in the minimum of the free energy. Moreover, we apply a Fourier transform $a_i^\dagger=\frac{1}{\sqrt{L}}\sum_k \ue^{i k r_i} a_k^\dagger$, leading finally to
%\eqa{
%H&=& \sum_k \Bigl[ a_k^\dagger a_k \bigl( 2\cdot 2S\sin\theta \cos^2\gamma \, \tilde\gamma_k^{(\alpha)}+2\cos\theta\cos\gamma \nonumber \\
%& &\phantom{\sum_k + a_k^\dagger a_k} -4\cdot 2S \sin\theta\sin^2\gamma\, \tilde\gamma_0^{(\alpha)}\bigr)  \nonumber \\
%& & \phantom{\sum_k} + \bigl(a_k^\dagger a_{-k}^\dagger + a_k a_{-k} \bigr) 2 S \sin\theta \cos^2\gamma \, \tilde\gamma_k^{(\alpha)} \Bigr]\\
%& & + 2S \sin\theta\cos^2\gamma \sum_k \tilde\gamma_k^{(\alpha)} +L (2S)^2 \sin\theta\sin^2\gamma \tilde\gamma_0^{(\alpha)} \nonumber \\ 
%& &- L 2S \cos\theta\cos\gamma  \nonumber
%}
\begin{widetext}
\eqa{
\label{eq:Hak}
H&=& \sum_k \Bigl[ a_k^\dagger a_k \,2 \bigl( 2S\sin\theta \cos^2\gamma \, \tilde\gamma_k^{(\alpha)}+\cos\theta\cos\gamma -4S \sin\theta\sin^2\gamma \tilde\gamma_0^{(\alpha)}\bigr)  
+ \bigl(a_k^\dagger a_{-k}^\dagger + a_k a_{-k} \bigr) 2 S \sin\theta \cos^2\gamma \, \tilde\gamma_k^{(\alpha)} \Bigr] \nonumber\\
& & + 2S \sin\theta\cos^2\gamma \sum_k \tilde\gamma_k^{(\alpha)} +L (2S)^2 \sin\theta\sin^2\gamma\, \tilde\gamma_0^{(\alpha)} 
- L 2S \cos\theta\cos\gamma  \,,
}
\end{widetext}
where we defined
\eq{
\tilde{\gamma}_k^{(\alpha)} = \sum_{\delta>0} \frac{(-1)^\delta}{\delta^\alpha} \cos k\delta\,.
}
This last abbreviation encorporates the entire long-range nature of the system, preserving the extreme simplicity and elegance of LSWT. 

Hamiltonian \eqref{eq:Hak} can now be diagonalized as usual by a Bogolioubov transformation, $a_k=\cosh\beta_k \, \alpha_k +\sinh\beta_k\, \alpha_{-k}^\dagger$, 
$a_{-k}^{\dagger}=\sinh\beta_k \, \alpha_k +\cosh\beta_k\, \alpha_{-k}^\dagger$. 
Demanding that the $\alpha_k$ obey bosonic commutation relations, and that only terms proportional to $\alpha_k^\dagger \alpha_k$ yield a contribution to $H$, one obtains the Bogolioubov angles $\cosh 2\beta_k=B_k/\omega_k$, $\sinh 2\beta_k=-2A_k/\omega_k$, and 
\begin{widetext}
\eq{
\label{eq:HBog}
H=\sum_k \omega_k \left(\alpha_k^\dagger \alpha_k +\frac 1 2\right) +\sum_k \left( A_k-\frac 1 2 B_k - 2S \cos\theta\cos\gamma +(2S)^2 \sin\theta\sin^2\gamma \, \tilde\gamma_0^{(\alpha)}  \right)\,,
}
where 
\begin{subequations}
\begin{eqnarray}
B_k&=&4S\sin\theta\cos^2\gamma \, \tilde\gamma_k^{(\alpha)} +2\cos\theta\cos\gamma - 8S\sin\theta\sin^2\gamma \, \tilde\gamma_0^{(\alpha)} \,, \\
A_k &= & 2S \sin\theta \cos^2\gamma \, \tilde\gamma_k^{(\alpha)} \,, 
\end{eqnarray}
\end{subequations}
\end{widetext}
and with dispersion relation 
\eq{
\label{eq:dispersionrelation}
\omega_k =\sqrt{B_k^2-4A_K^2}\,.
}
The ground state of Hamiltonian \eqref{eq:HBog}, $\ket{\psi_\mathrm{GS}}$, is found as the vacuum of Bogolioubov particles, $\alpha_k\ket{\Omega}=0\,\,\forall k$. 
We can now determine the free-energy minimum $\gamma$ by minimizing $\bra{\psi_\mathrm{GS}}H\ket{\psi_\mathrm{GS}}$. 
Alternatively, one can demand that terms that are linear in the spin-wave operators vanish at the energy minimum, which gives a condition on $\gamma$ as a function of $\theta$. 
For the case of $\theta=\pi/20$ that we use in the main text, we find $\gamma=0$ independent of $\alpha$, and the spins are strongly polarized in negative $z$ direction. 

\subsection{Spin-wave group velocity}

\begin{figure}
\centering
\includegraphics[width=0.5\columnwidth]{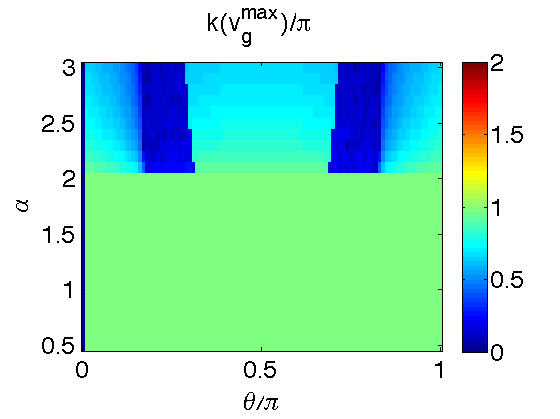}\includegraphics[width=0.5\columnwidth]{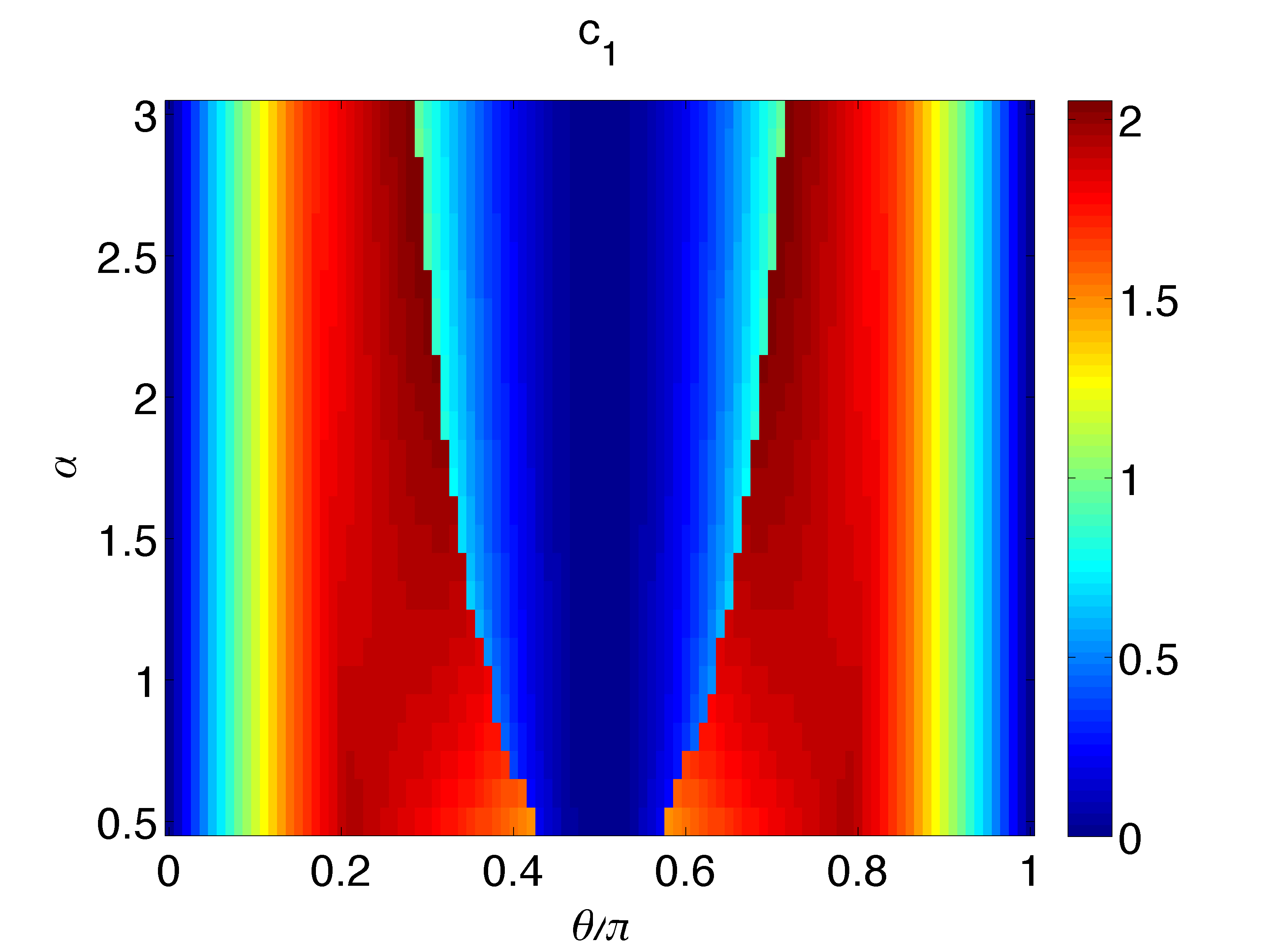}
\caption{ (Color online) 
{\bf Properties of the maximal group velocity.} 
{\bf Left:} 
The $k$ value at which $v_{\rm max}$ is found makes a transition to $k=\pi$ at $\alpha=2$. 
Data is for $L=500$. 
{\bf Right:} Where the constant $c_1$ vanishes, the system is dispersionless. This happens at the points $\theta=0,\pi/2,\pi$, where there are no non-commuting interactions in the Hamiltonian. Then, the system dynamics is localized. 
 }
\label{fig:k-of-vmax}
\end{figure}

The dispersion relation \eqref{eq:dispersionrelation} determines the group velocity, which is of the form
\eq{
\label{eq:vg_general}
v_g\equiv \frac{\partial \omega_k}{\partial k} = \frac{c_1 \frac{\partial \tilde{\gamma}_k^{(\alpha)} } {\partial k} } {\sqrt{ c_2 \tilde{\gamma}_k^{(\alpha)} + c_3 }} \,,
}
with $c_{1,2,3}$ constants. 
Divergences of $v_g$ (as given in the main text, see Fig.~\ref{fig:dispersionrelations_analytical}) can hence be found easily by analyzing $\tilde{\gamma}_k^{(\alpha)}$ and 
\eq{
\frac{\partial \tilde{\gamma}_k^{(\alpha)} } {\partial k} = -\sum_{\delta>0} \frac{(-1)^\delta}{\delta^{\alpha-1}} \sin k\delta\,.
}
As illustrated in the main text for the example of $\theta=\pi/20$  (Fig.~\ref{fig:dispersionrelations_analytical}b), there are two transitions that can be found generically through an analysis of $v_{\rm max}=\max_k v_g(k)$ as a function of $\alpha$. 
For $\alpha>2$, $v_{\rm max}$ is almost constant as a function of $\alpha$, whereas below it, it rises rather steeply with decreasing $\alpha$. 
This indicates a transition in the dynamical behavior of the system. 
Additionally, for $\alpha\gg 2$ the $k$-value where $v_{\mathrm{max}}$ is achieved lies around $k=\pi/2$ and changes slowly with decreasing $\alpha$, whereas at $\alpha=2$, it transitions to $k=\pi$ (Fig.~\ref{fig:k-of-vmax}, left panel). 

In the range $1<\alpha<2$, although $v_{\mathrm{max}}$ achieves large values, $v_{\mathrm{max}}/L$ scales to zero with increasing system size, indicating the locality of information in these systems. 
This changes drastically at $\alpha\leq1$, where $v_{\mathrm{max}}/L$ \emph{increases} with system size (see discussion in the main text). In that regime, the fastest mode reaches the boundaries earlier for larger systems; the information is distributed essentially instantaneously over the entire chain. 
Actually, the number of modes diverges for which the time to reach the boundary decreases with system size.
Consequently, $\forall t_{0}>0$, there exists a chain length $L_0$ such that $\forall L>L_0$ the number of $k$-modes with $v_g(k)>(L/2)/t_{0}$ is larger than any given $n_{0}\equiv c L^{(1-\alpha)/(3-\alpha)}$, with $c$ a constant.  For $\alpha<1$, $n_0$ diverges with $L$. 
In other words,  we find a diverging number of quasi-particle modes that reach the boundary before the (arbitrarily small) time $t_{0}$. 
The physical reason is simple: the mode spacing decreases faster with $L$ than the $k$-interval where $v_g(k)>(L/2)/t_{0}$ (see symbols in the main panel of Fig.~\ref{fig:dispersionrelations_analytical}a). 

This behavior is generically true, except when in Eq.~\eqref{eq:vg_general} we have $c_1 = 8S \sin\theta\cos^3\gamma (\cos\theta\cos\gamma - 4 S \sin\theta\sin^2\gamma \tilde{\gamma}_0^{(\alpha)} ) = 0$. 
As seen in Fig.~\ref{fig:k-of-vmax}, right panel, this happens only at $\theta=0,\pi/2,\pi,\dots$. At these parameter values, one has either only the magnetic field or only the spin--spin interactions, i.e., there are no non-commuting terms in the Hamiltonian. 
In that case, independent of divergences of $\frac{\partial \tilde{\gamma}_k^{(\alpha)} } {\partial k} $, the quasi-particle dispersion relation~\eqref{eq:dispersionrelation} becomes dispersionless, and the system dynamics is localized.

\subsection{Linear spin-wave theory and dynamics}

To evaluate the time evolution under $H$, we make use of the fact that all involved states remain Gaussian at all times. 
Since Hamiltonian \eqref{eq:Hak} is quadratic in the boson operators, its ground state is completely determined by the correlators 
\begin{subequations}
\begin{align}
F_{ij}&\equiv \bra{\psi_\mathrm{GS}}{a_i^\dagger a_j}\ket{\psi_\mathrm{GS}} +\frac 1 2 \delta_{ij}=\frac 1 {2 L} \sum_k \frac{B_k}{\omega_k} \ue^{i k (r_j-r_i)} \\
G_{ij}&=\bra{\psi_\mathrm{GS}}{a_i^\dagger a_j^\dagger}\ket{\psi_\mathrm{GS}}=\frac 1 {2 L} \sum_k \frac{-2A_k}{\omega_k} \ue^{i k (r_j-r_i)} 
\end{align}
\end{subequations}

At time $t=0$, we quench the system with a spin flip at site $m$, corresponding in LSWT to the operator $S_m^{x\prime}=\frac 1 2 (a_m + a_m^\dagger)$, and the state becomes $\ket{\psi_0}=S_m^{x\prime}\ket{\psi_\mathrm{GS}}/{\sqrt{\mathcal N}}$, where ${\mathcal N} = \left[ \Re \left(\bra{\psi_\mathrm{GS}}a_m a_m\ket{\psi_\mathrm{GS}}\right)  + \bra{\psi_\mathrm{GS}}a_m^\dagger a_m\ket{\psi_\mathrm{GS}}  + \frac{1}{2}\right]/2$ is the normalization 
\footnote{After applying $S_m^{x\prime}$ to the bosonic ground states, the occupation number at site $m$ can acquire values larger than unity, which are un-physical in the initial spin Hamiltonian. However, under the time evolution, the density spreads very fast far below 1, so that the mapping between spin-waves and bosons is restored. A possibility to avoid this problem would be to apply only a partial spin flip. 
}. 
Since the initial state is Gaussian, expectation values after the quench such as $\braket{a_i^\dagger a_j}(t=0)\equiv \bra{\psi_\mathrm{GS}}{S_m^{x\prime} a_i^\dagger a_j S_m^{x\prime}}\ket{\psi_\mathrm{GS}}/{{\mathcal N}}$ can be decomposed into a combination of expectation values of two-point correlations before the quench, using Wick's theorem \cite{Fetter1971}.
For example, 
$\bra{\psi_\mathrm{GS}} a_m^\dagger a_i^\dagger a_j a_m^\dagger \ket{\psi_\mathrm{GS}} =
\bra{\psi_\mathrm{GS}} a_m^\dagger a_i^\dagger \ket{\psi_\mathrm{GS}}  \bra{\psi_\mathrm{GS}} a_j a_m^\dagger \ket{\psi_\mathrm{GS}} 
+\bra{\psi_\mathrm{GS}} a_m^\dagger a_j \ket{\psi_\mathrm{GS}}  \bra{\psi_\mathrm{GS}} a_i^\dagger a_m^\dagger \ket{\psi_\mathrm{GS}} 
+\bra{\psi_\mathrm{GS}} a_m^\dagger a_m^\dagger \ket{\psi_\mathrm{GS}}  \bra{\psi_\mathrm{GS}} a_i^\dagger a_j  \ket{\psi_\mathrm{GS}}
$.
Since the correlations after the quench remain Gaussian, and since a Gaussian state remains Gaussian under the application of a quadratic Hamiltonian, it is sufficient to consider the time evolution of the two-point correlators. 

\begin{widetext}
To compute the time evolution, we use Heisenberg's equation of motion, which for an arbitrary operator $\mathcal A$ reads in the small-time limit, 
${\mathcal A}(t+\Delta t) = {\mathcal A}(t) + {i \Delta t} \left[ H, {\mathcal  A}\right]$. 
We find 
\begin{subequations}
\begin{align}
F_{ab}(t+\Delta t) =&  
F_{ab}
+ {i\Delta t}\, 2S \sin\theta\cos^2\gamma \Bigl[ 
\sum_{i\neq a} \frac{(-1)^{i-a}}{\left|i-a\right|^\alpha} \bigl( G_{ib}^\star + F_{ib} \bigr)
- \sum_{i\neq b} \frac{(-1)^{i-b}}{\left|i-b\right|^\alpha} \bigl( G_{ai} +F_{ai} \bigr)  
\Bigr]
\\
G_{ab}(t+\Delta t) = & G_{ab}
+ {i\Delta t}\Bigl\{ 2S \sin\theta\cos^2\gamma \Bigl[ 
\sum_{i\neq a} \frac{(-1)^{i-a}}{\left|i-a\right|^\alpha} \bigl( G_{ib} + F_{bi} \bigr)
+ \sum_{i\neq b} \frac{(-1)^{i-b}}{\left|i-b\right|^\alpha} \bigl( G_{ia} + F_{ai}  \bigr)  
\Bigr] \nonumber \\
& \qquad\qquad \quad \quad + 4\bigl( \cos\theta\cos\gamma - 4S \sin\theta\sin^2\gamma \, \tilde\gamma_0^{(\alpha)} \bigr) G_{ab} \Bigr\} 
\end{align}
\end{subequations}
where the right hand side is to be evaluated at time $t$. 
In our numerical evaluation, we chose $\Delta t=0.002$, and checked that a further decrease does not improve the results. 
In the main text, we plot as a function of time the deviation of the magnetization from $-1/2$, $\delta m_i\equiv\braket{S_i^{z}}+1/2$, which is nothing else than $F_{ii}(t)-1/2$. 

\end{widetext}

\section{Power-law interactions are reproducing if and only if $\alpha>1$}

Typical Lieb--Robinson bounds with the associated velocity are defined for short-range interacting systems, i.e., interactions that decay at least exponentially with distance $i-j$ between lattice sites $i$ and $j$. 
For interactions $K(i-j)$ that decay slower than exponential, one can still define Lieb--Robinson bounds on the commutators of operators --- similar to the bound \eqref{eq:lbr}, but without a constant Lieb--Robinson velocity --- as long as $K(i-j)$ is reproducing \cite{Hastings2010a}, i.e., if it fulfills the condition
\eq{
\label{eq:reproducing}
\sum_{m=-\frac L 2..\frac L 2,m\neq i,j}  K(i-m) K(m-j) \leq \lambda K(i-j) \quad \forall\,i,j
}
for some constant $\lambda$. For simplicity, we discuss here open boundary condition for a chain of length $L+1$ with $L$ even. 
For a power law $K(i-j)=1 /\left|i-j\right|^{\alpha}$, this condition is fulfilled if the decay is faster than $\alpha\geq 1$ and violated for $\alpha<1$, as we will show now. 

\subsection{Power-law interactions are reproducing if $\alpha>1$}
It is convenient to rewrite the condition \eqref{eq:reproducing} using the definition 
\eq{
P(i,j)\equiv \sum_{m=-\frac L 2..\frac L 2,m\neq i,j} \frac{\left|i-j\right|^{\alpha} }{ \left|i-m\right|^{\alpha} \left|m-j\right|^{\alpha}}\,,
}
so that it becomes $P(i,j)\leq \lambda$.
To show that $1/\left|i-j\right|^{\alpha}$ is reproducing for $\alpha\geq 1$, we need to demonstrate that $P(i,j)$ converges with $L$ for any $i$, $j$. 
Consider $i$ and $j$ placed symmetrically at positions $\pm\delta/2$ for $\delta$ even. 
Then, 
\eqa{
\label{eq:Pupperbound}
P\left(-\frac \delta 2,\frac \delta 2\right)&=&\sum_{|m|>\frac \delta 2} \frac {\delta^{\alpha} }{(m+\frac \delta 2)^{\alpha} (m-\frac \delta 2)^{\alpha}}\nonumber\\
&+&\sum_{|m|<\frac \delta 2}\frac {\delta^{\alpha} }{(\frac \delta 2+m)^{\alpha} (\frac \delta 2-m)^{\alpha}} \,.
}

Let us treat the two sums separately. 
The first sum is upper bounded by $\sum_{|m|>\delta/2} \frac {\delta^{\alpha} }{(m-\delta/2)^{2\alpha}} $, the last term of which reads $M \equiv \frac{4^\alpha \delta^\alpha}{(L-\delta)^{2\alpha}}$. For constant $\delta$, this goes to zero as $\propto L^{-2\alpha}$. 
From this, it would seem that this sum converges for $\alpha>1/2$. 
However, one can consider a more demanding scenario, which shows that convergence is reached only for  $\alpha>1$, namely, if one lets $\delta$ increase with system size, $\delta=\delta(L)=c L^\beta$. Here, the condition $\delta<L$ demands $\beta\leq 1$. 
Then, 
\eq{
M=\frac{4^\alpha c^\alpha L^{\alpha\beta}}{L^{2\alpha}(1-cL^{\beta-1})^{2\alpha}} 
\leq \frac{4^\alpha c^\alpha}{L^{\alpha}(1-\epsilon)^{2\alpha}}\,,
}
where we used $\beta\leq 1$ to bound $L^{\alpha\beta}\leq L^\alpha$ and $cL^{\beta-1}\leq \epsilon$ (with $\epsilon$ arbitrarily small for $\beta<1$, provided $L$ is sufficiently large, and $\epsilon=c<1$ for $\beta=1$).
Therefore, $M$ decays at least as fast as $L^{-\alpha}$, meaning that the sum over it is assured to converge for $\alpha>1$. 

For constant $\delta$, the second sum in Eq.~\eqref{eq:Pupperbound} is constant. The only way it can increase is by increasing $\delta$ in some way with $L$. To study if this can make it diverge, consider the difference of when it is evaluated at $\delta$ and $\delta+2$, 
\begin{align}
&	\sum_{|m|<\frac \delta 2 +1}\frac {(\delta+2)^{\alpha} }{(\frac \delta 2 +1+m)^{\alpha} (\frac \delta 2+1-m)^{\alpha}} \nonumber \\
-& \sum_{|m|<\frac \delta 2}\frac {\delta^{\alpha} }{(\frac \delta 2+m)^{\alpha} (\frac \delta 2-m)^{\alpha}} \nonumber \\
=& 2 \sum_{0<m<\frac \delta 2}\frac 1 {(\frac \delta 2-m)^{\alpha}} \left[\frac {(\delta+2)^{\alpha} }{(\frac \delta 2 +m+2)^{\alpha} } - \frac {\delta^{\alpha} }{(\frac \delta 2+m)^{\alpha}}\right] \nonumber \\
+& \frac{(\delta+2)^\alpha}{(\frac \delta 2 +1)^{2\alpha}} - \frac{\delta^\alpha}{(\frac \delta 2)^{2\alpha}} \\
+& \sum_{m=\pm1} \frac {(\delta+2)^{\alpha} }{(\frac \delta 2+1 +m)^{\alpha} (\frac \delta 2+1-m)^{\alpha}}\,, \nonumber
\end{align}
where we regrouped the terms of the two sums into contributions from the outermost summands, the one at the origin, and the additional summands that are inserted beside the origin upon increasing $\delta$. 
One can show for the first sum of this expression that all terms are negative, as is the case for the contribution at the origin. Now, we only have to show that the last few terms decay sufficiently fast. In fact, they decay as $\delta^{-\alpha}$, so that even increasing $\delta$ proportional to $L$ leads to a convergent sum as long as $\alpha>1$. Therefore, in this case, also the second sum in Eq.~\eqref{eq:Pupperbound} converges. 
The argumentation here can be carried over to positions deviating from the symmetric case $i,j=\pm\delta/2$. 
We have thus demonstrated that $K(i-j)\propto 1/|i-j|^\alpha$ is reproducing for $\alpha>1$. 

\subsection{Power-law interactions are non-reproducing if $\alpha\leq1$}
To show that $K(i-j)$ is non-reproducing for $\alpha\leq 1$, it is sufficient to demonstrate the divergence of $P(i,j)$ for a specific case, which can easily be done for the choice $i=-L/2$, $j=L/2$, 
\eqa{
P(-L/2,L/2)&=&\sum_{m=-L/2+1}^{L/2-1} \frac {L^{\alpha} }{(m+L/2)^{\alpha} (m-L/2)^{\alpha}}  \nonumber \\
& = & \sum_{m=1}^{L-1}  \frac{1} {m^{\alpha} (1-\frac{m}{L})^{\alpha}}\geq \sum_{m=1}^{L-1} \frac{1} {m^{\alpha}}
}
The last sum converges towards the Riemann--zeta function $\zeta(\alpha)$. This lower bound for $K(i-j)$, therefore, diverges for $\alpha\leq 1$, where $K(i-j)$ is hence non-reproducing. 
This property explains the violation of the Lieb--Robinson bounds for small $\alpha$.

\end{document}